\documentclass{css2020}

\usepackage{epsfig}

\newcommand{\PACS}{\MSC}

\title{Astrophysical Bounds on Mirror Dark Matter, Derived from Binary Pulsars Timing Data}
 
\author{Itzhak Goldman$^{1,2}$  \\
\vskip 2mm {\small
$^1$
Department  of Physics, Afeka Engineering College\\
Tel Aviv, Israel\\
goldman@afeka.ac.il \\
$^2$
Department of Astrophysics, Tel Aviv University \\
Tel Aviv, Israel}}

\abstract{  
Mirror Dark Matter (MDM) has been considered as an elegant framework for a particle theory of Dark Matter (DM).
It is supposed that there exists a dark sector which is  mirror of the ordinary matter. Some  MDM models   
 allow  particle interactions mirror and ordinary matter, in addition to the gravitational interaction.  The possibility of neutron to mirror neutron transition has recently been discussed both from theoretical and experimental 
perspectives. 
This paper is based on a previous work in which we obtained stringent upper limits on the possibility of converting neutrons to mirror neutrons in the interiors of neutron stars, by using timing data of binary pulsars. 
 Such a transition would imply mass loss in neutron stars  leading to a significant change of orbital
period of neutron star binary systems. The observational  bounds on the period changes of such binaries,
therefore put strong limits on the above transition rate and hence on the neutron -- mirror-neutron mixing parameter $\epsilon'$.   
    Our limits are  much stronger
than the values required to explain the neutron decay anomaly via $n-n'$ mixing.
}



\begin{document}

\maketitle

\section{Introduction} 
 
 The idea of mirror dark matter (MDM) has been considered by various authors \cite{mirror1}.  
 In this class of models each ordinary particle $p$ has a mirror particle $p'$ counterpart. 
   There is,    a generic problem   in that the extra  mirror neutrinos ($\nu'$) and mirror photon ($\gamma'$)  contribute  too much to the number of degrees of freedom at the Big Bang Nucleosynthesis (BBN) epoch, destroying the success of the big bang nucleosynthesis predictions. A possible way out is to  assume a breaking of the $Z_2$ symmetry in the early universe so as to have asymmetric inflationary reheating in the two sectors (ordinary matter and MDM) resulting in a lower reheat temperature ($T'$)  of the mirror sector compared to  (~$T$) of the visible one~\cite{asym}. This breaking eventually trickles down to the low energies leading in general to a splitting the mirror and visible fermion masses~\cite{MN}. The above mentioned symmetric picture could however remain almost
exact if the asymmetric inflation picture is carefully chosen. Cosmologies  of such scenarios have been discussed in ~\cite {foot} and references therein

.  
 
 An interesting new phenomenon is possible in almost exact mirror models, if there are interactions mixing the neutron with the mirror neutron state ( denoted by $\epsilon_{n-n'}\equiv \epsilon'$).  In such a case, one can expect $n\to n'$ oscillations  to take place in the laboratory~\cite{BB} and indeed there are ongoing and already completed searches for such oscillations~\cite{yuri} at various neutron facilities.

 We note that  $n-n'$ oscillation is   similar to neutron-anti-neutron oscillation suggested very early~\cite{nnbar} and  extensively  discussed  in the literature;  For recent reviews, see~\cite{review}.   The rate governing the $n-\bar n$ oscillations was limited by laboratory experiments to be
\begin{eqnarray}
\epsilon_{n \bar{n}}= 1/{\tau_{n\bar{n}}} < 10^{-8}~ Sec^{-1} ~{\rm or}~ 10^{-23}~{\rm eV}
\end{eqnarray}
 
 We focus on the possibility that  $n\to n'$ transitions can lead to mass loss of a neutron star and if the latter   is a member of a binary
s system, then this mass loss   affects the binary period. Indeed~\cite{GN}  the   mass loss of any kind of such neutron stars implies an increase of the
orbital period  of the binary system $P_b$.  In the following sections we find that observational  limits on 
 $|\dot{P}_b/P_b|$ of binary pulsars  
 yield stringent bounds  on the $n-n'$ mixing parameter $\epsilon'$.  
  
\section{Transition of a   neutron star to a mixed neutron-mirror neutron star induced by $n-n'$ mixing }

We note  that $n\to n'$  transitions,
kinematically forbidden in nuclei, can occur in neutron stars.
  The neutrons in neutron stars  are mainly bound by gravity and \emph{ not} by
nuclear forces.  Let
us, suppose that an  $n\to n'$
conversion occurred at some point  in the star. Under the pressure   a neighboring
neutron then will be pushed to "hole"  generated
by the converted neutron,  gaining in the process kinetic energy  
 which is of  order of the Fermi energy $E_F$. Additional  energy  gain is obtained since 
  the produced $n'$ gravitates to the center of the star and a
surface neutron replaces the neutron which went into the above mentioned  "hole".
Therefore the net  (eventually radiated via neutrino and mirror neutrino emission)  energy stemming from an early,  single $nn'$ transition  
 is: 
 
 \begin{equation}
  m_n c^2\left (e^{\phi (R)}  - e^{ \phi(0)}\right)  +<E_F>   
 \end{equation}

where $e^{\phi(r)} = g_{oo}^{1/2} (r)$,   $e^{\phi(R)} = ( 1- 2 GM/c^2) ^{1/2}
$  and $<E_F>$ is the average Fermi energy of the disappeared neutron. 
Thus the gravitational mass of the neutron star will decrease.  This has led \cite {mana} to propose that $(n\to n')$ oscillations   generating   completely mixed $nn'$ stars, may   explain   
the observed mass distribution of neutron stars.

\section{The rate of $n \rightarrow n'$  transition}

We express the
  $n\rightarrow n'$ transition rate $\Gamma( n\rightarrow{n'})$ by
 
 \begin{equation} 
  \Gamma( n\rightarrow{n'}) = 
\Gamma(nn)P_{nn'}   
 \end{equation}
where $\Gamma_{nn}$ is the rate of $nn$ collisions  and $P_{nn'}$ is the probability of having $n'$ (rather than   $n$) at the time of the   collision.
 
  Eq.(3)  is based  on the assumption  ~\cite{cowsik} in which
one assumes that:

a)  The coherent buildup of the  $|n'>$ component in the initial purely $|n>$ state
of the two component system, proceeds unimpeded by nuclear interactions
during the time of flight between two consecutive collisions.

b) The coherent build-up stops upon collision and the $n'$ part is
released as out-going mirror neutron particles.

Denoting by $t_{nn} $ the mean free time for neutron neutron collisions and by $\epsilon'$ the rate for $n$ to $n'$
oscillation,  the Hamiltonian in the two dimensional $|n>,|n'>$ Hilbert space leads to
$$P_{nn'} =[\epsilon'\cdot  t_{nn}]^2)$$

Substituting the above $P_{nn'}$ and $\Gamma_{nn} = t_{nn}^{-1} $ in Eq. (3)
one gets  
\begin{eqnarray}
    \Gamma_{n\to n'} = t_{nn}{\epsilon'}^2       \label{new}
    \end{eqnarray}
Taking $ t_{nn}\approx {10^{-23}} ~sec$, the flight time of a neutron a 
$O(Fermi)$ distance  at a speed   $\sim 1/3 c$  yields
\begin {eqnarray}
\Gamma_{n\to{n'}} =  6\times 10^{-8} [\epsilon'/{10^{-11} eV}]^2~{yr}^{-1}
\end{eqnarray}

 In what follows we shall use astrophysical data to obtain bounds on $\Gamma_{n\to{n'}} $ and employ  
Eq. (5) to derive bounds on $\epsilon'$. 

\section{Neutron star models and their descendant fully mixed neutron - mirror neutron stars}

To obtain  the resulting mass and radius decrease, we solved numerically  the TOV equations with a commonly used equation of state \cite{Steiner+2010} .  We first solve the  TOV  structure equations for pure neutrons. Then for the same   baryon number we solve the TOV equations for a totally mixed neutron - mirror neutron star. We employ the method used   by us in \cite{GMNRT 2013}. In this specific case we have two   fluids obeying the same nuclear  equation of state where the only interaction is through the gravitational field.

We considered 3 different models each characterized by its total baryon number (that does not change in  the transition).
For each baryon mass we calculated the initial and final mass and radius. We also calculated the initial 
$e^{\phi(R)} - e^{\Phi( 0) }$. The resulting models are:

\begin{itemize}
\item[\underline{\bf  $M_b= 1.81 M_{\odot}$}] 

$$  \ \ \ M= 1.57 M_{\odot}, \ \ \ \ R=12.2 km, \ \ \  e^{\phi(R)} - e^{\Phi(0 ) }=0.15  $$

$$  \ \ \ M_{nn'}= 1.43 M_{\odot}= 0.91  M\ \     \ \ \ \ R_{nn'}=8.8 km= 0.72 R  $$

\item[\underline{\bf $M_b= 2.17 M_{\odot}$ }] 

$$  \ \ \ M= 1.82 M_{\odot} \ \ \ \ R=12.37 km,  \ \ \  e^{\phi(R)} - e^{\Phi( 0) }=0.17   $$

$$  M_{nn'}= 1.62 M_{\odot}= 0.89   M\ \ \ \ R_{nn'}=8.8 km= 0.71 R,  \ \ \  $$

\item[\underline{\bf $M_b= 2.44 M_{\odot}$} ]

$$ \ \ M= 1.97 M_{\odot},  \ \ \ \ R=12.4,  \ \ \    e^{\phi(R)} - e^{\Phi(0 ) }=0.19 $$

$$  M_{nn'}= 1.7 M_{\odot}= 0.86   M\ \ \ \ R_{nn'}=8.65 km= 0.7  R ,  \ \ \ $$
 
\end{itemize} 

\section{The relation between the mixing parameter $\epsilon'$ and the mass loss rate}

Returning to the main goal of the paper, namely limiting   $ \epsilon'$, we need to  relate  
  the rate of neutron to mirror neutron transition  to the  stellar $\frac{\dot M}{M} $.
 Since neutrons from the entire volume of the neutron star can transform to $n'$, the resulting  total mass loss of the neutron star due to this is proportional  to $\Gamma_{n\to n'}$.

A great advantage is that   all the different   binary pulsars (with variety of masses,   spin down ages and companions ) should conform to a single fundamental parameter: $\epsilon'$.  Thus, we can use the youngest pulsars to set stringent limits on $\epsilon'$. In turn, this implies that also the older pulsars are still in  the process of transition. In particular, for small enough values of $\epsilon'$ even these older pulsars may be at the very initial stages of the pure to mixed star transition.

We next present two estimates of  $\frac{\dot M}{M} $. 

{\bf First Estimate}

 The first estimate uses   the average value $\frac{\dot M}{M}  \approx    \Gamma_{nn'}\frac{\Delta M}{M  }$ during the complete  transition to  a mixed star where,  $\Delta M $ is the total  mass reduction. 
The   $\frac{\Delta M}{M}= 0.09\div 0.14$ obtained in the previous section  then yields a representative value

\begin{equation}
\left|\frac{\dot M}{M}\right|\approx 0.12 \Gamma_{nn'}
\end{equation}

{\bf Second Estimate}
 
 Here we use the results of the numerical solutions of the neutron stars   presented in  section 4.
 We apply  Eq. (2) to the very beginning of the transition process and find

 \begin{equation}
\left | \frac{\dot M}{M} \right| \geq   \left ( e^and{\phi (R)} - e^{\phi (0)} \right )    \Gamma_{nn'}=  ( 0.15\div 0.19) \Gamma_{nn'} 
 \end{equation}

It reassuring that the estimates by the two methods agree up to a factor of 1.5. It also makes sense that at the very beginning the process is  faster than that derived from the first method which represents a time average over the entire transition.

Thus we adopt 
\begin{equation}
\left | \frac{\dot M}{M} \right| \geq   0.14\Gamma_{nn'}
 \end{equation}
as a representing value during the entire transition episode.

\section{Limits on  $\epsilon'$ derived from timing observations of binary pulsars}

  Jeans (1924) \cite{Jeans}, pointed out that   the mass of the star that emits electromagnetic radiation decreases with time and therefore 
the orbital elements of a binary system should evolve with time. Assuming that in the local  frame of each star the radiation 
emission is  spherically symmetric, he obtained

\begin{equation}
M a =\ {\rm constant}
\end{equation} 

where $a$ is the semi-major axis and $M=m_1 + m_2$ is the total mass of the system.
This and the  expression for the binary period

\begin{equation}
P_b = 2\pi\sqrt{\frac{a^3}{G M}}
\end{equation}

imply

\begin{equation}
\frac{\dot{P}_b}{P_b}=  -2 \frac{ \dot M}{M}
\end{equation}  
  Since $ \dot M <0 $,  $ \dot{P}_b >0 $ so that  the orbital period keeps increasing.
   
Except for the nature of emission process this is the same as the situation addressed here.
We consider next four binary pulsars and obtain the the limits on the present rate of the mass change for each system.

In general, only neutron stars which are still in the process of transiting to a mixed star will exhibit mass loss.   A priory,  for extremely "high" $\Gamma_{nn'}$ some of the older observed  pulsars in binaries may be "too old".

 {\bf PSR 1916+13}

We use the data of \cite{Weisberg etal 2010}. 
   The spin down age of the pulsar is  $1.1\times 10^8 yr $. It   is commonly assumed      that the pulsar companion is a neutron star, so that there is no mass transfer between the binary members. Also one neglects    the   mass accretion from the interstellar medium.  After accounting for the expected gravitational radiation,  Galactic acceleration, and further dynamical corrections there is still some limited room for a positive change of binary period that could have followed from the mass decrease. 
\begin{equation}
 \frac{\dot P_b}{P_b} < 1.36\times 10^{-11} yr^{-1} 
\end{equation}

implying
$$\left|\frac{\dot M}{M}\right| <6.8\times 10^{-12} yr^{-1}$$
for the two  neutron stars losing mass.  Using Eqs. (5)   and   (14) the bound
  
  $$\epsilon'< 2.8 \times 10^{-13} eV$$
  follows.

 {\bf PSR J1141-6545}

This is a young pulsar of age about $2\times 10^6 yr$. The companion is a massive white dwarf of mass   $0.98  M_{\odot}$ and the neutron star has a mass $1.3  M_{\odot}$ \cite{verbiest, verbiest2} 

This system is a superb laboratory for testing GR. Using the data from the above observational papers one finds that the residual (subtracting from the measured value the gravitational radiation terms as well as the galactic acceleration and the kinematic effect and allowing for the uncertainties of all the above) positive possible value of the orbital period rateThe  mirror neutrinos ($\nu'$) and mirror photon ($\gamma'$)  contribute  too much to the number of degrees of freedom at the Big Bang Nucleosynthesis (BBN of change is very small

  $$\frac{\dot P_b}{P_b} < 1.84\times 10^{-12} yr^{-1}$$
implying for the given masses and taking into account that only the neutron star is undergoing  the mass loss

$$\left|\frac{\dot M}{M}\right|= \frac{M +M_c}{2M} \frac{\dot P_b}{P_b} < 1.6 \times 10^{-12} yr^{-1 }$$
where $M_c$ is the mass of the white dwarf companion.

This implies a limit

$$\epsilon'< 1.4 \times 10^{-13} eV$$

This result is very important, as for the  other three pulsars pulsars, one may have argued that the the transition $n\to n'$ has already finished. This young pulsar closes the door on this argument.

  {\bf PSR J0437-4715}

This is a neutron star- white dwarf binary with masses of $1.76 M_{\odot}$ and $0.25 M_{\odot}$, respectively.  The spin-down age is $ 1.6 \times 10^9 \ yr$. Using the data from the above observational papers one finds that the residual
 (subtracting from the measured value the gravitational radiation terms as well as the galactic acceleration and the kinematic effect and allowing for the uncertainties of all the above) positive possible value of the orbital period rate of change is
\cite{verbiest2}
$$\frac{\dot P_b}{P_b} < 2.8\times 10^{-11} yr^{-1}$$

implying

$$\left| \frac{\dot M}{M}\right|  <1.6 \times 10^{-11} yr^{-1}$$

and therefore
 
 $$\epsilon'< 4.4 \times 10^{-13} eV$$
 
{\bf PSR J1952+2630 }
 
This pulsar is in a binary orbit with a $(0.93\div1.4) M_\odot$ white dwarf companion~\cite{Lazarus et al. 2014}  .  
 Its spin down age is $7.7 \times 10^7$ yr, the orbital period is $0.39$ days and during 800 days of follow-up the error on the period is $ 7\times 10^{-12 }$ days. This leads to $\frac{\dot P_b}{P_b}< 8.2 \times 10^{-12} yr^{-1}$. 

Taking into account that only the pulsar losses mass   we get
$$\left| \frac{\dot M}{M}\right| = \left(\frac{M +M_c}{2M}\right) \frac{\dot P_b}{P_b} < 7 \times 10^{-12} yr^{-1 }$$ 
where $M_c$ is the mass of the white dwarf companion.

Thus one finds
  $$\epsilon'< 2.9 \times 10^{-13} eV$$

The    expected period change due to gravitational radiation in this pulsar is 2 orders of magnitude smaller, and thus does not interfere with the derived limit.

\section{Discussion} In this paper, used  astrophysical data, notably, precision 
pulsar timing measurements  to 
 strongly constrain a putative $n\to n'$ tThe  mirror neutrinos ($\nu'$) and mirror photon ($\gamma'$)  contribute  too much to the number of degrees of freedom at the Big Bang Nucleosynthesis (BBNransition in neutron stars.
 
We solved numerically the geneandral relativistic structure equations for neutron stars with three different baryon mass. Using a realistic nuclear matter equation of stat, we first solved for a pure ordinary neutron star and then (for the same baryon mass) we obtained the solution for a fully mixed neutron- mirror neutron star. In this way we found the average mass reduction rate over the the time span of the transition.  From the solution of the ordinary neutron star we found the mass reduction at the very beginning of the process. We found that   two methods yield   similar relations between the rate of mass change of the star and the rate of the microscopic  process.

This allows us to restrict   the mixing parameter between. 
We find that the key parameter $\epsilon'$ responsible for $n\to n'$ transition
  is restricted to be below $4. 4 \times 10^{-13}$ eV. 
  
 All four binary pulsars  considered  here yield quite similar limits  on  $\epsilon'$ .  Of particular importance is the the limits are quite the same even though   systems ages vary between  $2\times 10^6\ yr$ and $1.6 \times 10^9  yr$. The presence of the youngest binary pulsar refutes the possibility that for the older pulsars the process has been already finished.
 
Our limits on $\epsilon'$ also exclude the possibility that $n n'$ oscillations can explain the neutron decay anomaly
 \cite{bere3}

\section*{Acknowledgments}
This work has been supported by the Afeka College Research Fund.

\end{document}